\documentclass{aa}  

\usepackage{graphicx}
\usepackage{txfonts}
\newcommand{\pv}{\ensuremath{P_V}}
\newcommand{\nv}{\ensuremath{N_V}}

\newcommand{\bz}{\ensuremath{\langle B_z \rangle}}

\newcommand{\bs}{\ensuremath{\langle \vert B \vert \rangle}}

\newcommand{\te}{\ensuremath{T_{\rm eff}}}
\newcommand{\esp}{ESPaDOnS}
\newcommand{\wda}{WD\,1105--340}
\newcommand{\wdb}{WD\,2150+591}
\newcommand{\ha}{H$\alpha$}

\begin{document}

   \title{Discovery of kilogauss magnetic fields on the nearby white
   dwarfs \wda\ and \wdb 
   \thanks{Based on observations obtained at the
   Canada-France-Hawaii Telescope (CFHT) which is operated by the
   National Research Council of Canada, the Institut National des
   Sciences de l'Univers of the Centre National de la Recherche
   Scientifique of France, and the University of Hawaii, under
   Programme 18AC006; on observations made with ESO Telescopes at the
   La Silla Paranal Observatory, under programme IDs 0101.D-0103 and
   0102.D-0045(A),; and at the Observatorios de Canarias del IAC with
   the William Herschel Telescope, operated on the island of La Palma
   by the Isaac Newton Group of Telescopes in the Observatorio del
   Roque de los Muchachos, under Programme 18B-P15.}  }

   \author{J. D Landstreet
          \inst{1,2}
          \and
          S. Bagnulo
	  \inst{1}
          }

   \institute{Armagh Observatory and Planetarium, College Hill, 
              Armagh, BT61 9DG, United Kingdom\\
              \email{John.Landstreet@Armagh.ac.uk, Stefano.Bagnulo@Armagh.ac.uk}
         \and
             University of Western Ontario, Department of Physics 
             \& Astronomy, London, Ontario N6G 1P7, Canada\\
             \email{jlandstr@uwo.ca}
             }

   \date{Received November 13, 2018; accepted January 29, 2019}

  \abstract{ Magnetic fields are present in roughly 10\% of white
  dwarfs. These fields affect the structure and evolution of such
  stars, and may provide clues about their earlier evolution
  history. Particularly important for statistical studies is the
  collection of high-precision spectropolarimetric observations of (1)
  complete magnitude-limited samples and (2) complete volume-limited
  samples of white dwarfs. In the course of one of our surveys we have
  discovered previously unknown kG-level magnetic fields on two nearby
  white dwarfs, WD\,1105--340 and WD\,2150+591. Both stars are
  brighter than $m_V = 15$. WD\,2150+591 is within the 20-pc volume
  around the Sun, while WD\,1105--340 is just beyond 25\,pc in
  distance. These discoveries increase the small sample of such
  weak-field white dwarfs from 21 to 23 stars.  Our data appear consistent
  with roughly dipolar field topology, but it also appears that the
  surface field structure may be more complex on the older star than
  on the younger one,  a result similar to one found earlier in
  our study of the weak-field stars WD\,2034+372 and WD\,2359--434. 
  This encourages further efforts to uncover a clear link between
  magnetic morphology and stellar evolution.}

   \keywords{stars: magnetic fields --
                stars: individual --
                polarisation --
                white dwarfs 
               }

   \titlerunning{Kilogauss magnetic fields on two nearby white dwarfs}

   \maketitle
%

\section{Introduction}

Magnetic fields play important roles in stars. They transfer angular
momentum, both internally during stellar evolution, and externally
during periods of accretion or mass loss. Even a fairly weak magnetic
field can suppress convection in stellar atmospheres and affect
cooling times of extremely old white dwarfs \citep{Tremetal15}. From
the properties of upper main sequence magnetic stars, it is clear that
a field can strongly alter the surface chemistry of a star. In cool
stars with dynamo fields, magnetism can lead to quite spectacular
surface activity, as is seen in the Sun. 

Magnetic fields were first discovered on white dwarfs (WDs) almost 50
years ago \citep[e.g.][]{Kempetal70,AngeLand71,LandAnge71}. They have
been studied fairly extensively since that time
\citep{Ange78,Land92,Ferretal15}. For about 30 years, new magnetic
white dwarfs (MWDs) were discovered at a rate of about 1--2 per year,
at first from detection of broad-band circular polarisation, and later
also from observation of spectral lines strongly shifted and distorted
by the Zeeman effect and its extension to high magnetic field
strengths.  

More recently, many hundreds of new MWDs have been
discovered thanks to the Sloan Digital Sky Survey
\citep[e.g.][]{Schmetal03,Kepletal16}, but because of the low
resolving power and modest S/N of thes SDSS spectra, the new
discoveries are almost entirely found in the field strength range
between about 1--2 to 80\,MG. At the same time, interest in possible
weaker fields has led to a small number of discoveries of fields
below 1\,MG, down to a few kG
\citep[e.g.][]{SchmSmit94,Aznaetal04,Koesetal09,KawkVenn14}. It is 
now clear that the range of field strengths in MWDs covers at least
five dex, from a few kG to nearly 1000\,MG.

Some of the fields are observed to vary. When the variations are
studied, they appear to be periodic, and apparently arise only from
the rotation of the underlying star, with periods usually of the
order of hours or days. Both single spectra and series of
spectropolarimeteric observations have been modelled in a few cases,
and it appears that the field structure is organised at a large scale,
sometimes with a configuration simple enough to be described as a
dipole tilted at an angle with respect to the rotation axis.  MWD
fields appear to be examples of ``fossil fields'', retained in the
stars from prior evolution stages, which are no longer being actively
generated by stellar dynamo action.

Evolution of magnetic fields leading to MWDs is not understood. Even
evolution during the WD stage is still unclear. Samples of MWDs that
might illuminate evolution during the WD phase are frequently
magnitude-limited samples in which one can study the overall frequency
of MWDs, and (in principle) the evolution of frequencies. Such samples
are heavily biased by V magnitude, and require uncertain corrections
for incompleteness.  To study evolution during the WD stage, it is
better to use volume-limited samples. In such samples, older stars can
be seen as descendants of younger stars, so we can almost directly
study magnetic field evolution: do the older stars in the sample show
a similar fraction of MWDs as the younger stars? Are the fields found
on old MWDs weaker than on young MWDs, or stronger? Do the field
structures evolve with age in a systematic way?

We are trying to answer some of these questions by achieving a largely
complete magnetic survey of both a reasonable magnitude-limited sample
down to somewhere between $m_V = 14$ and 15, and of the 20-pc
volume. The 20-pc sample in particular requires including as many of
the coolest nearby WDs as possible. Many of the cooler (and fainter)
WDs within 20 pc of the Sun have never been searched for magnetic
fields weaker than 1--2 MG, the limit reached with normal
low-dispersion classification spectroscopy of DA and DB stars. Some
nearby WDs (particularly featureless DCs and DQs with only molecular
band absorption) still have almost no useful limit on possible
magnetic fields. 

We also aim to study the magnetic field structure of
a usefully large sample of WDs with fields in the kG range.

In the course of our surveys we have discovered two bright new
weak-field magnetic white dwarfs. The field of one was not detected
with classification spectroscopy, and the other is a newly recognised
white dwarf. These two stars are discussed in this paper.

\section{Observations}

Weak fields can be detected in primarily two ways, both using the
changes to spectral lines resulting from the Zeeman effect, which
splits and polarises spectral lines when a magnetic field is present
on the surface of the star \citep[e.g.][]{DonaLand09,BagnLand15}. The
effect on the normal flux or intensity spectrum (Stokes $I$) is that
single spectra lines, such as those of the Balmer series of H, split
into multiple lines. In the case of the Balmer lines, each single line
core splits into three components, one of which (for fields below
about 1\,MG) remains at nearly the zero-field position, while two
flanking components appear, one at shorter and one at longer
wavelengths, separated from the central component by a wavelength
difference that grows with the strength of the magnetic field. For
\ha\ the two flanking ($\sigma$) components are shifted from
the central ($\pi$) component by about 1\,\AA\ per 50\,kG of field
strength. More generally, the splitting depends on wavelength and
field strength as 
\begin{equation} 
\Delta \lambda_{\rm Z} = 4.67\times10^{-13} g \lambda_0^2 \bs,
\end{equation}
where $\Delta \lambda_{\rm Z}$ is the separation between the $\pi$ and
one $\sigma$ component, $\lambda_0$ is the wavelength of the spectral
line in the absence of a magnetic field, \bs\ is the magnetic field
strength averaged over the visible stellar hemisphere, and  the
effective Land\'{e} factor $g$ is a constant for each line, equal to
1.0 for Balmer lines.

The detectability of such splitting depends on both the resolution and
the S/N ratio of the WD spectrum. The low-resolution ($R \sim 800$)
usually used to classify and characterise WDs can only resolve
magnetic splitting for fields of 1--2\,MG in the blue, or for fields
of a few hundred kG at \ha. Further, with a S/N of less than about 20
the Zeeman splitting, if present, may be mistaken for noise in the
line core. With higher resolution ($R > 8000$, say) the splitting
can be recognised at \ha\ for fields of about 100\,kG, and with a
resolving power of 50\,000 or more the splitting can be observed in
\ha\ down to the limit set by the intrinsic width of the central
non-LTE line core of about 1\,\AA, so that Zeeman splitting can be
resolved down to about 50\,kG and line core broadening can be detected
down to about 20\,kG. 

In addition to splitting a single spectral line into multiple (often
three) components, the Zeeman effect also leads to polarisation of the
various components. In particular, the $\sigma$ components are
circularised polarised in opposite senses when the field has a
substantial component along the line of sight, both in emission and in
absorption.  The mean wavelengths $\lambda_{\rm r}$ and
$\lambda_{\rm l}$ of a spectral line as viewed in right and left
circularly polarised light are different, and the separation
$\Delta \lambda_{\rm los}$ between these two mean wavelengths of a line is
proportional to the line-of-sight component of the magnetic field averaged 
over the visible stellar hemisphere, \bz: 
\begin{equation}
   \Delta \lambda_{\rm los} = \lambda_{\rm r} - \lambda_{\rm l} =
   2 \times 4.67 \times 10^{-13} g \lambda_0^2 \bz,
\end{equation}
where all wavelengths are measured in \AA.

The separation $\Delta \lambda_{\rm los}$ can be measured in two
somewhat different ways. (1) in the case where the field is
sufficiently weak (or even not present) that the shape of the observed
Stokes $I$ line profile is not perturbed significantly (at the
observational resolving power) from the non-magnetic shape, so that
both circularly polarised line profiles $I_{\rm r}$ and $I_{\rm l}$
are effectively the same shape, but centred on slightly different
wavelengths $\lambda_{\rm r}$ and $\lambda_{\rm l}$, the intensity
difference $V = I_{\rm r} - I_{\rm l}$ at each wavelength is
approximately given by $V = (1/2) \times dI/d\lambda \times \Delta
\lambda_{\rm los}$, where $I(\lambda) = I_{\rm r}(\lambda) + I_{\rm
l}(\lambda)$. Using Eqn.~(2), this leads directly to 
\begin{equation}
   V/I = -4.67 \times 10^{-13} g \lambda_0^2 (1/I)
          ({\rm d}I/{\rm d}\lambda) \bz.
\end{equation}
As all the quantites in this equation except \bz\ can be measured at
each point along the line profile, the many instances of this equation
can be considered as a linear least squares problem to be solved for
the slope \bz\ and its uncertainty by standard methods
\citep{Bagnetal15,BagnLand18}.

(2) In the more general case where the two line profiles $I_{\rm r}$
and $I_{\rm l}$ are significantly different because the Zeeman
splitting is large enough to show clearly in the observed profiles,
the separation $\Delta \lambda_{\rm los}$ is evaluated numerically. It
is easily shown that the expression to use is 
\begin{equation}
   \bz = -2.14\,10^{12} \frac{\int v V(v) {\rm d}v}
                      {g \lambda_0 c \int (I_{\rm cont} - I(v)){\rm d}v}
\end{equation}
where the intensity $I$ and circular polarisation $V$ are expressed as
function of velocity $v$ relative to the central wavelength of the
line $\lambda_0$, $I_{\rm cont}$ is the local continuum of the line being measured, $g = 1.0$, and $c$ is the speed of light
\citep{Math89,Donaetal97}. The numerical application of this
expression for WD fields is discussed at some length by 
\citet{Landetal15}. (It
should be noted that there is some uncertainty as to exactly where to
place the zero-point for the measurements of the core depth $I_{\rm
cont}$ that appears in Eqn.~(2), which affects somewhat the field
strength reported, but this does not affect the S/N of detection of a
value of \bz\ that is significantly different from zero, as both the
value of \bz\ and of its uncertainty are divided by the same
denominator.)

Because the two polarised spectra required to obtain such a difference
spectrum are obtained simultaneously through almost identical optical
trains, the difference spectrum can be used to detect very small
differences between the right and left circularly polarised
spectra. In practice, for bright WDs, Stokes $I$ and $V$ spectra can be
used to detect and measure magnetic fields \bz\ as small as 1\,kG
\citep{BagnLand18}.

\subsection{\wda}

\begin{figure}[ht]
\scalebox{0.37}{
\includegraphics[width=24cm,trim=0cm 13.5cm 1cm 1cm,clip]{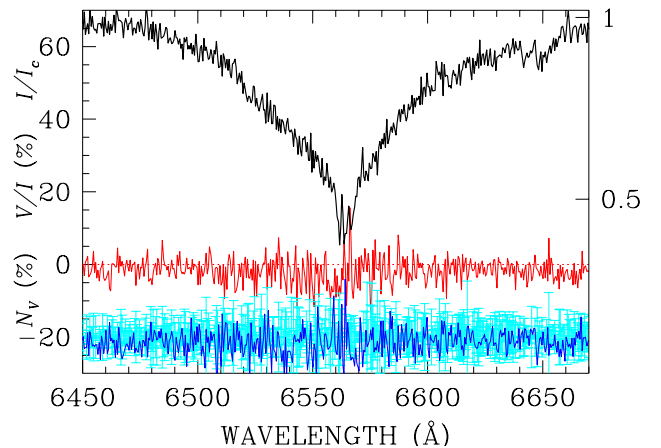}}\\
\caption{\label{Fig_wd11-esp} 
\ha\ in the polarised spectrum of WD1105--340 obtained with \esp. The $I$
flux spectrum (black) is normalised to 1.0 in the wings of \ha\ 
(the scale of the $I$ spectrum is shown on the right side of the box). The
circular polarisation $V/I$ spectrum (red) is shown in \%; the scale
is on the left side. Below the $\pv = V/I$ spectrum is the \nv\
spectrum, centred on $-25$\%, with error bars for both \pv\ and \nv\ in
light blue. The spectrum  (originally acquired with $R = 65000$)
has been binned in boxes 0.5\,\AA\ wide to reduce visual noise without
smearing the Zeeman pattern in the line core. }
\end{figure}

\wda\ was observed as part of an ongoing survey of relatively bright
($V < 15$) white dwarfs for which low resolution spectroscopy
\citep{Subaetal07} has shown that no field greater than about
1\,MG is present, but for which measurements sensitive to a kG-level
field have not been made. The basic parameters of this star are
listed in Table~\ref{Tab_obs-stars}; these are generally taken from
the Gaia Collaboration DR2 \citeyear{Gaiaetal16,Gaiaetal18}, from
\citet{Gianetal11}, and from \citet{Subaetal17}. This star was observed
with the high-resolution echelle spectropolarimeter
\esp\ on the Canada-France-Hawaii telescope, which produces a 
spectrum with a resolving power of $R \approx 65000$ over the range 
3800\,\AA\ to 1.04\,$\mu$m. Details of the
observation are given in Table~\ref{Tab_obs-log}. A clear Zeeman
triplet due to a magnetic field of $\bs \sim 120$\,kG was immediately
recognised in \ha, as shown in Figure~\ref{Fig_wd11-esp}. However,
although the circular polarisation spectrum was also observed for this
star, the S/N achieved was not sufficiently high to reveal Zeeman
polarisation in any of the Balmer lines.

\begin{figure}
\scalebox{0.33}{
\includegraphics*[width=28cm,trim=0.5cm 13cm 0cm 1cm,clip]{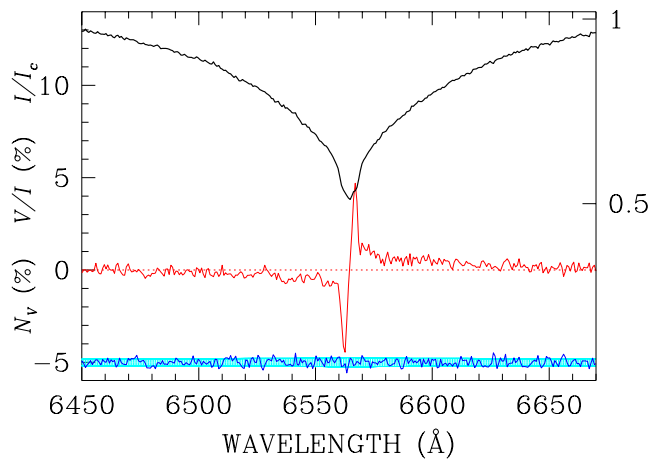}}\\
\caption{\label{Fig_wd11-fors} 
The \ha\ polarised spectrum of WD1105--340 observed with FORS. The
scale of $I$ is shown on the right side of the panel. The scale of
$\pv = V/I$, shown on the left side in \%, is centred on 0.0\%, while
the scale of the null spectrum $\nv$ is centred on --5\%. The broad
blue band shows the uncertainties in
\pv\ and \nv.  }
\end{figure}

\begin{figure}
\scalebox{0.33}{
\includegraphics*[width=28cm,trim=0.5cm 2cm 0cm 0cm,clip]{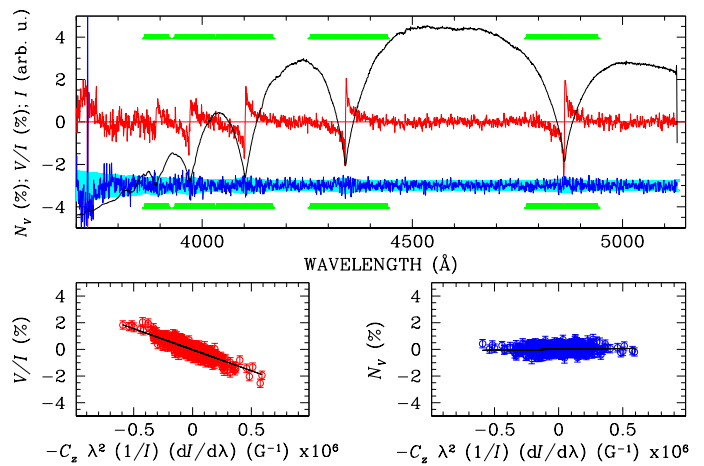}}\\
\caption{\label{Fig_wd11-fors-blue} 
The blue polarised spectrum of WD1105--340 observed with FORS. The
scale of $I$ is arbitrary.  The meaning of symbols in the upper
panel is the same as in Fig.~\ref{Fig_wd11-fors}. The green bars
above and below the spectra show the wavelength windows included in
the computation of \bz. The lower left panel shows the correlation of
Stokes $V/I$ with the local slope of the spectral line, while the
lower right panel shows the same correlation for the null spectrum
$\nv$. }
\end{figure}

This observation was followed up with one observation of \ha\ and
a later observation of the Balmer lines H$\beta$ through H8 using the
FORS spectropolarimeter of the European Southern Observatory's 8-m
Antu telescope (see Table~\ref{Tab_obs-log}).  The settings used
on FORS yield a spectrum with resolving power of $R \approx 3600$ over
the range 5800 to 7300\,\AA, and a spectrum in the range 3670 to
5130\,\AA\ with $R \approx 2500$. The resolving power of FORS was not
high enough to show clear Zeeman splitting, but it is clear from the
intensity spectrum that the core of \ha\ is quite abnormal The
higher Balmer line cores appear fairly normal. The larger collecting
area of the 8-m telescope made possible high $S/N$ detection of a
strong signal around each of the Balmer line cores in the
circular polarisation spectra, clearly confirming the presence of a
magnetic field on this star. These two spectra are illustrated in
Figs~\ref{Fig_wd11-fors} and
\ref{Fig_wd11-fors-blue}.

Detailed analysis of these three spectra will be discussed in
Section~3.1.

\subsection{\wdb}

This object was identified as a probable new white dwarf member of the
20-pc volume around the Sun by \citet{Holletal18} from a careful
examination of the Gaia DR2 release. This star is labelled by Hollands
et al. as 215139.93+591734.6(J2015.5); it is also identified as Gaia
DR2 2202703050401536000. We will designate it as WD\,2150+591.  
The parameters of the WD were taken from the Gaia Collaboration DR2 
and from \citet{Holletal18}. The cooling age was estimated by 
interpolating in Table~1 of \citet{Bergetal95}.

We observed \wdb\ during an observing run with the ISIS
spectropolarimeter of the William Herschel Telescope at the
Observatorio del Roque de los Muchachos. This run was devoted
primarily to searching for kG-level fields on stars of the 20-pc
volume for which the available field strength limits are of the order
of 1\,MG. During this run, we observed simultaneously with the blue
arm ($R \approx 4100$, spectral range 3530 -- 5370\,\AA) and with the
red arm ($R \approx 10500$, spectral range 6000 -- 7035\,\AA) of
ISIS. Details of the observations of this star are given in
Table~\ref{Tab_obs-log}. During the first observation, the shape of
the (extremely weak) \ha\ line core showed clear Zeeman splitting
characterised by a relatively sharp $\pi$ component and quite broad
$\sigma$ components. However, the circular polarisation spectrum did
not yield a clear detection of a magnetic signature. No higher Balmer
line components are detected in our spectra of this DA star of $\te
\approx 5100$\,K, which is close to the temperature limit below which
no Balmer lines are visible in the spectra of white dwarfs with H-rich
atmospheres. No metal lines appear to be present in either the
blue or the red spectra of \wdb; there is no evidence from the optical
spectrum connecting this very cool DA WD to the very interesting and
growing class of cool DAZ stars with weak magnetic fields, such as
NLTT\,7547 \citep{Kawketal19}. 

\begin{figure}
\scalebox{0.38}{
\includegraphics*[width=24cm,trim=0cm 1cm 0cm 0cm,clip]{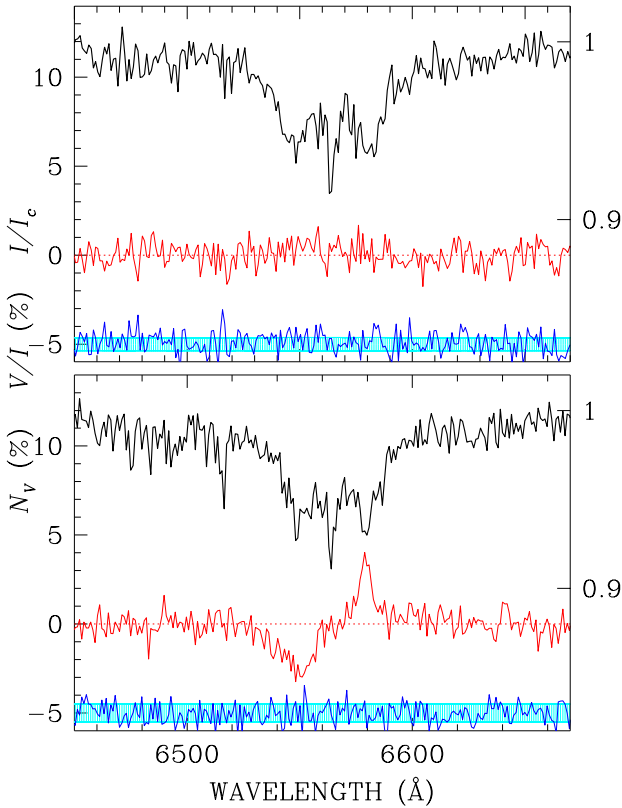}}\\
\caption{\label{Fig_g21-isis} 
The two polarised \ha\ spectra of WD\,2150+591 obtained with ISIS (19
Sept: upper panel, 22 Sept: lower panel), binned in
1\,\AA\ bins to reduce noise.  The meaning of the symbols is the
same as in Fig.~2.  }
\end{figure}

Because of the detection of the apparent Zeeman splitting in the first
spectrum of this star, it was re-observed on two subsequent nights of
the ISIS run. Due to an error in decker selection, the second spectrum
lacks a background measurement, and cannot be measured reliably. The
third spectrum showed the Zeeman triplet previously seen, and also
yielded a detection of strong circular polarisation in the two
$\sigma$ components. Details of the three observations are reported in
Table~\ref{Tab_obs-log}. The two useable spectra are illustrated in
Figure~\ref{Fig_g21-isis}. Measurement of the spectra will be
described in Section~3.2. 


\begin{table}
\caption{Parameters of new magnetic white dwarfs}
\label{Tab_obs-stars}
\centering
\begin{tabular}{l r r}
\hline\hline
                  & WD\,1105--340 & WD\,2150+591       \\
\hline    
$\alpha$ (J2000)  & 11 07 47.90  & 21 51 39.93  \\  
$\delta$ (J2000)  & -34 20 51.49 & +59 17 34.6  \\
$\pi$ (mas)       & 38.22        & 118.12       \\
Johnson $V$       & 13.66        &              \\
Gaia $G$          & 13.72        & 14.39        \\
Spectrum          &  DA\,3.6     &  DA\,9.9     \\
$T_{\rm eff}$ (K) &  13970       &  5095        \\
$\log g$ (cgs)    &  8.05        &  7.98        \\
age (Gyr)         &  0.31        &  5.0         \\
\hline
\end{tabular}
\end{table}

\begin{table*}
\caption{Observations of newly discovered magnetic white dwarfs}
\label{Tab_obs-log}
\centering
\begin{tabular}{l l l l l r r@{$\pm$}l r@{$\pm $}l}
\hline\hline
Star          & Instrument & Grism/     &  MJD        &\multicolumn{1}{c}{Date} & Exp.  &\multicolumn{2}{c}{\bz}&\multicolumn{2}{c}{\bs}\\
              &            & Grating    &             &  (yr-mo-da UT)    &\multicolumn{1}{c}{(s)}&\multicolumn{2}{c}{(kG)}&\multicolumn{2}{c}{(kG)}\\
\hline
WD\,1105--340 & \esp       &            &  58156.510  & 2018-02-07 12:14  &  2268  &$  -9.0$&6.2& 129 &  5  \\
              & FORS       &  1200R     &  58238.094  & 2018-04-30 02:15  &  5400  &$ -30.3$&1.0& 160 & 15  \\
              & FORS       &  1200B     &  58468.293  & 2018-12-16 07:02  &  3307  &$ -31.5$&0.6&\multicolumn{2}{c}{}  \\
WD\,2150+591  & ISIS       &  R1200R    &  58380.972  & 2018-09-19 23:20  &  3600  &$ +30  $& 11& 820 & 20  \\
              & ISIS 
  &  R1200R    &  58383.019  & 2018-09-22 00:28  &  3600  &\multicolumn{2}{c}{}&\multicolumn{2}{c}{} \\
              & ISIS 
  &  R1200R    &  58383.956  & 2018-09-22 22:56  &  3600  &$ -260$&15  & 700 & 20  \\

\hline
\end{tabular}
\end{table*}

\section{Measurement of magnetic fields}


\subsection{WD1105--340}

As discussed in the previous section, we can detect and measure both
\bs\ and \bz\ on \wda. The \esp\ spectrum of this star shows a clear 
Zeeman triplet in \ha\ when the data are binned in windows 0.5\,\AA\
wide. We can estimate the mean field modulus using Eqn.~(1). We have
measured the $\pi - \sigma$ separation both with eye estimates of the
wavelength positions of the three components, and by fitting Gaussians
to each component to measure its position. The two estimates yield
essentially the same separation of $2.56 \pm 0.1$\,\AA, implying a
field strength of $130$\,kG. This value means that only eight of the
600 or more MWDs with estimated field strengths 
\citep{Ferretal15} are known to have well-confirmed fields weaker
than \wda.

The \ha\ FORS spectrum of \wda\ has been obtained with a relatively low
spectral resolution of around $R \approx 3600$. With this resolving
power, the resolution broadening of the core of \ha\ is comparable to
the separation of the Zeeman components. Thus the unusual width of the
core is clear in the $I$ spectrum, but the components are not
separated. We have carried out some simple numerical \ha\ line core
modelling, using the spectrum synthesis program {\sc zeeman.f} as
described by \citet{Landetal17}, and find that the model line core has
the same width as the FORS line core for a field of $\bs \approx 160
\pm 15$\,kG.  

The blue FORS spectrum of \wda\ does not display enough
broadening to estimate \bs.

The initial polarised \esp\ spectrum of \wda\ showed a clear Zeeman
pattern in $I$, but the S/N ratio of the Stokes $V$ signal was not high
enough to provide a secure detection of Zeeman polarisation. However,
the observed signal still allows us to measure the mean line-of-sight
component of the field averaged over the visible hemisphere, \bz. 

Using Eqn.~4 on the \esp\ spectrum of \wda, we find a value
of \bz\ that is not significantly different from zero, so this measure
does not provide confirming evidence of the presence of a magnetic
field on \wda. However, it does provide quite a strong constraint on
the \bz\ field strength at the moment of observation. 

The low-resolution red FORS polarised spectrum of \wda, as noted above,
does not permit a very precise measurement of \bs, but it does provide
a very convincing detection of the longitudinal magnetic field, with
peak values of the polarisation $V/I$ of about $\pm 4$\,\%, detected
with more than 15$\sigma$ significance in the characteristic S-shaped
polarisation signal. When this polarisation signal is used to
determine the value of \bz\ using the correlation method based on
Eqn.~(3), as discussed in Sec.~2, the deduced field strength is $-20.5
\pm 1$ kG. This value is probably an underestimate of the magnitude of
the actual \bz\ field value, as the splitting in \ha\ is large enough
that the underlying assumption of a nearly unperturbed line core is no
longer very accurate.

We have also derived a value of \bz\ from the \ha\ FORS spectrum by
measuring the separation between the mean wavelength of the line core
as observed in right and left circularly polarised light, using the
same numerical method to evaluate Eqn.~(4) as was employed for
the observation with \esp. In this case, the field strength is found
to be $-30.3 \pm 1.0$\,kG, a somewhat larger absolute value than
derived using the correlation method, as expected. We adopt this value
as being very probably more accurate than that obtained with the
correlation technique.

In contrast, the Stokes $I$ profiles of the higher Balmer lines
of the blue FORS spectrum of \wda\ are perturbed little enough that
using the normal correlation method based on Eqn.~(3) is expected
to have acceptable accuracy. For this spectrum, the derived value of
\bz\ is $-31.5 \pm 0.6$\,kG.

The magnetic measurements of \wda\  are summarised in
Table~\ref{Tab_obs-log}.

The values of \bz\ derived from the three available spectra 
clearly indicate that the line-of-sight component of
the magnetic field of \wda\ is variable, with a currently unknown
period. Note this conclusion is not affected by the lack of detection
of a non-zero \bz\ field in the \esp\ spectrum, because all
measurements are sufficiently precise to clearly establish that they
are very different. Further observations are expected to provide a
rotation period for this star, and sufficient data to allow simple
modelling of the global structure of the field, as has been described
for the weak-field MWDs WD\,2047+372 and WD\,2359--434
\citep{Landetal17}.

\subsection{WD\,2150+591}

The temperature estimated for \wdb\ by \citet{Holletal18} 
is $\te = 5095$\,K. With
this effective temperature, only \ha\ is usefully detectable in the
spectrum. On this star \ha\ is a few percent deep, and is easily seen
to be split by the Zeeman effect due to a field of several hundred
kG. The mean field modulus may be estimated using a measurement of the
$\pi - \sigma$ separation based on eye estimates of the centroid
wavelength of each component, or by fitting Gaussians to each
component. Although the $\sigma$ components are quite broad, the
results of the the two methods are in good agreement, and lead to
separation estimates of 16.4 and 14.1\,\AA\ for the observations
on Sept 19 and 22, which in turn suggest  varying values of \bs\ ranging
at least between $\bs \approx 820$ and $700$\,kG.

The polarisation data from these observations can clearly not be
analysed successfully by the correlation method used for measurement
of very weak fields, the method used to analyse most of our previous
magnetic observations with ESO-FORS \citep{Bagnetal15}. This is because the
complete separation of the $\pi$ and $\sigma$ components means that
the strongest polarisation signals are correlated not with the slope
of the blended line components but with the individual depths of the
$\sigma$ components. However, analysis of the separation between line
centroids as seen in right and left circular polarisation, using
Eqn.~(4), works very well. This method reveals that on Sept~19 the
longitudinal field \bz\ was quite close to zero,
while on Sept 22 it was nearly $-260$\,kG.

It is clear from these data that the magnetic field of \wdb\ is very
securely detected, that it can be measured with considerable accuracy
using the spectropolarimeters available in the northern hemisphere, and
that the field is variable on a time scale of hours or days. Future
observations should provide the rotation period of this star, and
allow simple modelling of the geometry of the magnetic field
distribution over the stellar surface. 

The magnetic measurements of \wdb\ are summarised in Table~\ref{Tab_obs-log}.

\section{Preliminary modelling of field structure}

With only two or three observations for each of the two newly
discovered MWDs, we are not able to constrain the surface distribution
of magnetic flux over each star very tightly. However, we can begin
to extract even from such limited measurements some useful
information about possible models.

The basic simple model normally used as a first approximation for the
field structure of a MWD is the surface magnetic flux distribution
produced by a magnetic dipole at the centre of the star. The 
justification for using this model is that when we measure a value of
$|\bz|$ that is a substantial fraction (say, 20\% or more) of \bs, we
can deduce that the global structure of the stellar field is such that
there is one hemisphere that has most magnetic flux lines entering the
star, while on the other hemisphere most flux lines are emerging. In
the contrasting situation of a lot of small-scale magnetic structure,
as on the surface of the Sun, much local cancellation of field
contributions to \bz\ occurs, and $|\bz|$ is much smaller than \bs. A
centred dipole (or indeed a uniform field with all field lines through
the star parallel to one another) captures this tendency towards
relatively simple large-scale coherence with a small number of parameters. 

The required parameters of a dipole model are the surface field
strength on the magnetic axis at one of the poles $B_{\rm d}$, the
obliquity $\beta$ of the dipole axis to the rotation axis, and the
inclination $i$ of the rotation axis to the line of sight to the star.

If we limit ourselves to models of this class, the variability of \bz\
observed in the observations of each of the new MWDs immediately
provides real constraints. (1) It is clear that $i$ must be
significantly different from zero, for otherwise we would observe no
variations in \bz\ regardless of the values of $\beta$ and $B_{\rm
d}$.  (2) Similarly, $\beta$ must be substantially different from
zero, or again we would detect no variation in \bz\ independently of
the values of $i$ and $B_{\rm d}$. (3) We can estimate that $B_{\rm
d}$ will probably be at least 20 or 30\% larger than the largest value
of \bs\ observed. (4) Consider the sets of observations obtained for
each star. Unusually, one observation of each has the property that
\bz\ is very close to zero, while in the other observation(s) $|\bz| \sim
0.2
\bs$, as on \wda, or even $0.5 \bs$, as on \wdb. In the case of 
nearly zero \bz, the line of sight appears to point roughly towards
the magnetic equator, while in the case with a large value of $|\bz|$,
the line of sight points roughly toward a magnetic pole.

Thus for \wda, we arrive at a first guess model of a simple centred
dipole with both $i$ and $\beta$ quite different from zero, and
$B_{\rm d} \sim 150$\,kG. For \wdb\ our first guess model is also a
simple centred dipole with both angles large, and $B_{\rm d} \sim
900$\,kG. 

However, there is a striking difference between the appearance of
the \ha\ lines of \wda\ and \wdb\ that should provide additional
information about the field structure. For \wda,
the two $\sigma$ components of the resolved Zeeman pattern seen in the
\esp\ data (Fig.~\ref{Fig_wd11-esp}) are about the same width as the central
$\pi$ component, suggesting that the dispersion in
the local field strength $|B|$ over the visible hemisphere is small
compared to its mean value. In
contrast, the $\sigma$ components of \wdb\ are so broad
(Fig.~\ref{Fig_g21-isis}) that it is clear that the dispersion in
local $|B|$ is of the same order as the mean value.

In part this difference between \wda\ and \wdb\ may be due to the much
stronger field present on \wdb, which leads to significant variation
in the wavelengths of the $\sigma$ components due to even moderate
variations in the local value of $|B|$. However, this difference
between the two stars in the width of the $\sigma$ components relative
to the $\pi$ component is similar to the difference noted between
WD\,2047+372 (with $\bs = 60$\,kG and very well-defined $\sigma$
components) and WD\,2359--434 (with $\bs \sim 50 - 100$\,kG and very
diffuse $\sigma$ components) \citep{Landetal17}. For that pair of
stars, the different appearance of the $\sigma$ components was
apparently a symptom of rather different global magnetic field
structures, with that of the young star WD\,2047+372 being ``simpler''
than that of the considerably older WD\,2359--434. We suspect that this
difference may also be present on this pair of stars between the
relatively young \wda\ and the much older \wdb. 

With the limited data available, we will not try to refine these
modelling ideas further, but it is clear that if we can determine the
rotation period of each star, and obtain several polarised spectra
spread through the rotation period, we should be able to obtain
plausible, and much better constrained, models of the strength and
structure of the magnetic field over the stellar surface.

\section{Discussion and Conclusions}

In this paper we have reported the firm detection of sub-MG magnetic
fields on two relatively bright DA white dwarfs. \wda\ has a field
which, assuming that it can be at least roughly described as that of a
dipole with its axis at a non-zero angle to the rotation axis,
probably has a polar dipole field strength of order 150\,kG. \wdb\ may
have a very roughly dipolar field as well, with a polar 
field strength of the order of 900\,kG.

The discovery of these fields is significant for several
reasons. First, we have increased the small sample of 21 MWDs known to
have sub-MG fields by two, or by about 10\%. Since this range of field
strength covers almost half of the full known (logarithmic) field
strength range, these are clearly valuable additions to a rather
poorly known field strength domain.

Secondly, both the newly discovered kG-level MWDs are brighter than
$m_V = 15$. Only about ten of the previously known sub-MG MWDs are this
bright. The practical importance of this characteristic is that such
stars are far easier to study in detail, for example by identifying
rotation periods and mapping the surface magnetic field structure,
using available 4-m class telescopes, than the weak-field MWDs of
magnitude 16 or 17. And because both new MWDs are clearly magnetically
variable, they are both particularly well-suited to detailed study.

Thirdly, both stars are members of WD samples that are of substantial
interest. They both qualify as members of the magnitude-limited
samples that are often used for establishing the frequency of
occurrence of MWDs of various types. Thus the discovery of fields in
these two stars will change the best estimates of what fraction of WDs
have detectable fields, and of the distribution of field strengths over
the magnetic sub-sample. 

Furthermore, \wdb\ is also within the 20-pc sample of white dwarfs, the
best current proxy for studying observationally the time evolution of the
final stages of evolution of a fixed sample of stars in the Milky Way
\citep{Giametal12,Holbetal16,Holletal18}. 

Another significant feature of both these new MWDs (kindly
pointed out by the referee) is that both are members of wide binary
systems consisting of a white dwarf and a low-mass main sequence
star. As there is considerable interest currently in the possibility
that (some?) WD magnetism is a consequence of binary evolution through
a common envelope phase \citep{Toutetal08,Ferretal15,Brigetal18}, the
presence of weak fields in WD members of such binary systems is of
considerable interest. Note, however, that in both systems the current
separation of the WD and the M dwarf is of the order of 100\,AU, so
that the two stars may have effectively evolved this far as single
objects.

Finally, it appears that these two MWDs are consistent with the idea
that the surface field structures of young and old white dwarfs may
well be rather different from one another in terms of complexity. 
Remarkably, a similar contrast, between an apparently simple magnetic
field structure on the relatively young white dwarf WD\,2047+372 and a
more complex one on the older white dwarf WD\,2359--434, was reported
recently by \citet{Landetal17}.  This difference could turn out to be
an evolutionary effect that may be a useful clue about how a MWD field
evolves as the underlying star cools.

Thus both of these new MWDs are well worth monitoring for modelling
purposes, and we plan to obtain further observations in the near
future.

\begin{acknowledgements}

We thank the anonymous referee for several helpful suggestions and
improvements. JDL acknowledges the support of the Natural Sciences and
Engineering Research Council of Canada (NSERC), funding reference
number 6377-2016.

\end{acknowledgements}

%
%

\bibliography{sb-jdl}

\end{document}